\documentclass[10pt, letterpaper]{article}
\usepackage{amsfonts,amsmath,amsopn,amssymb,amsthm,bbm,latexsym,mathrsfs,pstricks,verbatim}
\usepackage{feynmf}
\usepackage[dvips]{graphicx
}

\numberwithin{equation}{section}

\long\def\symbolfootnote[#1]#2{\begingroup%
\def\thefootnote{\fnsymbol{footnote}}\footnote[#1]{#2}\endgroup}

\makeatletter
\def\underbracket{\@ifnextchar [
   {\@underbracket} {\@underbracket [\@bracketheight]}}
\def\@underbracket[#1]{\@ifnextchar [
   {\@under@bracket[#1]} {\@under@bracket[#1][0.2em]}}
\def\@under@bracket[#1][#2]#3{
   \mathop {\vtop {\m@th \ialign {##
            \crcr $\hfil \displaystyle {#3}\hfil$
            \crcr \noalign {\kern 3\p@ \nointerlineskip }\upbracketfill {#1}{#2}
            \crcr \noalign {\kern 3\p@ }
            \crcr }}}\limits}
\def\upbracketfill#1#2{$\m@th \setbox \z@ \hbox {$\braceld$}
                  \edef\@bracketheight{0.5pt}
                  \upbracketend{#1}{#2}
                  \leaders \vrule \@height #1 \@depth \z@ \hfill
                  \leaders \vrule \@height #1 \@depth \z@ \hfill
                  \upbracketend{#1}{#2}$}
\def\upbracketend#1#2{\vrule height #2 width #1\relax}

\def\overbracket{\@ifnextchar [
 {\@overbracket} {\@overbracket [\@bracketheight]}}
\def\@overbracket[#1]{\@ifnextchar [
 {\@over@bracket[#1]} {\@over@bracket[#1][0.2em]}}
\def\@over@bracket[#1][#2]#3{
\mathop {\vbox {\m@th \ialign {##
  \crcr \noalign {\kern 3\p@ }
  \downbracketfill {#1}{#2}
  \crcr \noalign {\kern 3\p@ \nointerlineskip }
  \crcr  $\hfil \displaystyle {#3}$
  \crcr
  } }}\limits}
\def\downbracketfill#1#2{$\m@th
  \setbox \z@ \vbox {$\braceld$}
  \edef\@bracketheight{0.5pt}
  \downbracketend{#1}{#2}
  \leaders \vrule \@height #1 \@depth \z@ \hfill
  \leaders \vrule \@height #1 \@depth \z@ \hfill
\downbracketend{#1}{#2}$}
\def\downbracketend#1#2{\vrule depth #2 width #1\relax}
\makeatother

\newcommand{\bra}[1]{\langle#1|}
\newcommand{\ket}[1]{|#1\rangle}

\begin{document}\setlength{\unitlength}{1mm}


\def\pplogo{\vbox{\kern-\headheight\kern -29pt
\halign{##&##\hfil\cr&{\ppnumber}\cr\rule{0pt}{2.5ex}&\ppdate\cr}}}
\makeatletter
\def\ps@firstpage{\ps@empty \def\@oddhead{\hss\pplogo}
  \let\@evenhead\@oddhead
}
\def\maketitle{\par
 \begingroup
 \def\thefootnote{\fnsymbol{footnote}}
 \def\@makefnmark{\hbox{$^{\@thefnmark}$\hss}}
 \if@twocolumn
 \twocolumn[\@maketitle]
 \else \newpage
 \global\@topnum\z@ \@maketitle \fi\thispagestyle{firstpage}\@thanks
 \endgroup
 \setcounter{footnote}{0}
 \let\maketitle\relax
 \let\@maketitle\relax
 \gdef\@thanks{}\gdef\@author{}\gdef\@title{}\let\thanks\relax}
\makeatother


\setcounter{page}0
\def\ppnumber{\vbox{\baselineskip14pt
}}
\def\ppdate{RUNHETC-2007-22} \date{}

\author{Jean-Fran\c{c}ois Fortin\footnote{jffor27@physics.rutgers.edu, JFF is supported by FQRNT.} \\
[7mm]
{\normalsize NHETC, Department of Physics and Astronomy,}\\
{\normalsize Rutgers University, Piscataway, NJ 08854, U.S.A.}\\}

\title{\bf \LARGE Spontaneously Broken Gauge Symmetry in SUSY Yang-Mills Theories with Matter} \maketitle \vskip 2cm

\vspace{-0.5cm}

\begin{abstract}
\noindent
The analysis of spontaneous gauge symmetry breaking of $N=1$ supersymmetric $SU(N_c)$ Yang-Mills theory with matter is performed.  The supersymmetric $R_\xi$-gauge is used and its non-local effects investigated.  Superpropagators and vertices are computed, and it is shown that the non-local terms introduced by the $R_\xi$-gauge-fixing are well-behaved in general gauge at one-loop.  It is argued that this feature generalizes to multiple loops.
\end{abstract}
\vspace{0.5cm}

\section{Introduction}\label{intro}

Although the extension of the $R_\xi$-gauge to supersymmetric gauge theories has been studied previously \cite{Ovrut:1981wa,Marcus:1983wb,Siegel:1995iu,Goldhaber:2004kn} confusion still remains about the results.  Ovrut and Wess \cite{Ovrut:1981wa} extended the $R_\xi$-gauge to spontaneously broken $SU(N_c)$ super Yang-Mills with matter and computed the superpropagators.  They did not however calculate the vertices nor did they stress the non-local behavior of these gauges.  Later Marcus, Sagnotti and Siegel \cite{Marcus:1983wb} and Siegel \cite{Siegel:1995iu} used equivalent gauge-fixing terms in the context of ten-dimensional Yang-Mills theory and four-dimensional $N=1$ superspace Gervais-Neveu gauge respectively.  Recently Goldhaber, Rebhan, van Nieuwenhuizen and Wimmer \cite{Goldhaber:2004kn} discussed the supersymmetric extension of the $R_\xi$-gauge and pointed out that non-local terms appear in the action.  They concluded that one might not be able to construct a local $R_\xi$-gauge theory in the superFeynman gauge due to the presence of non-local terms.

This paper re-introduces the $R_\xi$-gauge for supersymmetric Yang-Mills with matter and shows that the theory is well-defined at one-loop in general gauge.  The effectiveness of supersymmetric $R_\xi$-gauge relies on the projection operator for chiral fields \cite{Ovrut:1981wa} which however introduces non-local terms in the gauge-fixing term as described in \cite{Goldhaber:2004kn}.  These non-local terms show up in two different parts of the action, in the gauge-fixing action and the ghost action.  Most of these terms become gauge-dependent mass terms for quark and ghost superfields while non-zero vacuum expectation values give vector superfield mass term.  This is analogous to usual $R_\xi$-gauge \cite{'tHooft:1971rn} and Higgs mechanism \cite{Higgs:1964ia,Higgs:1964pj,Higgs:1966ev} in non-supersymmetric theories.  The non-local terms left are all of the same form and correspond to vertices between one quark superfield and two ghost superfields.  To one-loop, non-local contributions to the effective action are well-defined, the non-renormalization theorem of supersymmetric theories forcing several non-local diagrams to give a zero contribution.  Remaining contributions are mostly finite, the most divergent diagrams are only logarithmically divergent and do not require additional counterterms.

The massive vector superfields encountered in these theories are interesting on their own.  Indeed, supersymmetry (SUSY) could be the theory beyond the standard model of particles and it is not excluded that electroweak symmetry breaking happens at energy scales higher than SUSY breaking.  In this scenario massive vector superfields would be generated, which signals SUSY as the theory beyond the standard model.

The paper is constructed as follows.  In section \ref{SU} the $R_\xi$-gauge is introduced for $SU(N_c)$ super Yang-Mills theory with fundamental matter and the superpropagators and vertices in general gauge are computed.  In section \ref{oneloop} non-local contributions to the effective action at one-loop are computed from superFeynman diagrams.  The effects of these non-local terms are discussed and it is argued that higher-order corrections should have the same form.  Notation conventions follow \cite{Ovrut:1981wa} and are gathered in appendix \ref{notation} with other useful identities.  The computation of the ghost action is left for appendix \ref{ghost}.  Finally appendix \ref{propagatorvertex} lists the propagators and vertices in the GRS formalism \cite{Grisaru:1979wc}.  

\section{$SU(N_c)$ supersymmetric QCD with matter}\label{SU}

The starting point is $SU(N_c)$ supersymmetric QCD with $N_f$ flavors of quarks $Q_{in}$ in a given representation $R$ (in general complex, reducible) of the gauge group ($i,j=1,\ldots,N_f$ are flavor indices and $n,m=1,\ldots,\mbox{dim}\;R$ are gauge group index).  The gauge group representation $R$ carried by the quarks is chosen such that anomalies cancel.  The SUSY Yang-Mills action is given by
\begin{equation}
S_{\mbox{{\tiny inv}}}=\frac{1}{16g^2C_2(A)}\mbox{Tr}\left(\int d^6z\;W^\alpha W_\alpha+h.c.\right)+\int d^8z\;\overline{Q}_ie^VQ_i.
\end{equation}
The quarks $Q_{in}(z)$ are chiral superfields and the gauge bosons $V_{nm}(z)=V^a(z)T_{nm}^a$ are vector superfields.  Here $a,b=1,\ldots,N_c^2-1$ are indices in the adjoint representation $A$ of the gauge group and $T_{nm}^a$ are the generators of the gauge group in the representation $R$.  The generators of the vector superfield action are in the adjoint representation where $T(A)=C_2(A)$ (see appendix \ref{notation}) and the normalization is chosen such that the rescaling $V\rightarrow2gV$ leads to the canonical normalization.  For simplicity the superpotential is set to zero.  This avoids further complications due to additional propagators between (anti-)chiral quark superfields.  From the super field strength $W_\alpha=-\frac{1}{4}\bar{D}^2(e^{-V}D_\alpha e^V)$ the vector superfield action can be rewritten in a more convenient way as an integral over full superspace
\begin{eqnarray}
S_V &=& \frac{1}{16g^2C_2(A)}\mbox{Tr}\left(\int d^6z\;W^\alpha W_\alpha+h.c.\right)\nonumber\\
 &=& -\frac{1}{64g^2C_2(A)}\mbox{Tr}\left(\int d^6z\;\left(-\frac{\bar{D}^2}{4}\right)\left(\bar{D}^2(e^{-V}D^\alpha e^V)(e^{-V}D_\alpha e^V)\right)+h.c.\right)\\
 &=& -\frac{1}{64g^2C_2(A)}\mbox{Tr}\left(\int d^8z\;\bar{D}^2(e^{-V}D^\alpha e^V)(e^{-V}D_\alpha e^V)+h.c.\right).\nonumber
\end{eqnarray}
An interesting phenomenon is the SUSY analog \cite{Ovrut:1981wa} of the Higgs mechanism \cite{Higgs:1964ia,Higgs:1964pj,Higgs:1966ev} where quarks have non-zero vacuum expectation values.  These theories, which have massive vector superfields, could be relevant if e.g. electroweak symmetry breaking happens at higher energy scales than SUSY breaking.  With that in mind, the quark vacuum expectation values are chosen such that they do not break SUSY nor Poincar\'e invariance.  The simplest choice is
\begin{equation}
Q_{in}(z)=q_{in}+\Phi_{in}(z)
\end{equation}
where $q_{in}$ are constrained by the auxiliary field equations of motion.  The phenomenon giving rise to non-zero quark expectation values is not of interest here.  Expanding the action leads to
\begin{equation}
S_{\mbox{{\tiny inv}}}=-\frac{1}{64g^2C_2(A)}\mbox{Tr}\left(\int d^8z\;\bar{D}^2(e^{-V}D^\alpha e^V)(e^{-V}D_\alpha e^V)+h.c.\right)+\int d^8z\;(\overline{q}_i+\overline{\Phi}_i)e^V(q_i+\Phi_i).
\end{equation}
In order to cancel quark superfield/vector superfield cross-terms, one introduces the chiral gauge-fixing term
\begin{equation}\label{gaugefixing}
F^a=\bar{D}^2V^a+32g^2\xi\left(\frac{\bar{D}^2}{16\partial^2}\right)\overline{\Phi}_iT^aq_i
\end{equation}
which is, as component notation shows, the SUSY analog of the non-SUSY $R_\xi$-gauge.  $F^a$ is chosen chiral since gauge transformations have a chiral parameter $\Lambda^a$.  This choice of gauge-fixing term takes advantage of the chiral field projection operator $P_2=\frac{\bar{D}^2D^2}{16\partial^2}$ with $P_2\Phi(z)=\Phi(z)$.  However, as shown by the second term of equation (\ref{gaugefixing}), it forces the introduction of non-local terms in the action which can spoil the consistency of the theory in these gauges.  This non-locality will be studied more carefully in the following section.  The generating functional is gauge-fixed following the general procedure of the functional determinant
\begin{equation}
\Delta(V)=\int\mathbbm{D}\Lambda\mathbbm{D}\overline{\Lambda}\;\delta[F(V^{\Lambda,\overline{\Lambda}})-f]\delta[\overline{F}(V^{\Lambda,\overline{\Lambda}})-\overline{f}].
\end{equation}
Averaging over $f$ and $\overline{f}$ with a Gaussian weight factor results in the $SU(N_c)$ superQCD generating functional
\begin{eqnarray}
Z &=& \frac{1}{N}\int\mathbbm{D}f\mathbbm{D}\overline{f}\mathbbm{D}V\mathbbm{D}\Phi\mathbbm{D}\overline{\Phi}\mathbbm{D}\widetilde{\Phi}\mathbbm{D}\overline{\widetilde{\Phi}}\;\exp\left(-\frac{i}{32g^2\xi}\int d^8z\;\overline{f}^af^a\right)\Delta^{-1}(V)\Delta(V)e^{iS_{\mbox{{\tiny inv}}}}\nonumber\\
 &=& \int\mathbbm{D}V\mathbbm{D}\Phi\mathbbm{D}\overline{\Phi}\mathbbm{D}\widetilde{\Phi}\mathbbm{D}\overline{\widetilde{\Phi}}\mathbbm{D}c\mathbbm{D}\overline{c}\mathbbm{D}c'\mathbbm{D}\overline{c}'\;e^{iS_{\mbox{{\tiny inv}}}+iS_{\mbox{{\tiny GF}}}+iS_{\mbox{{\tiny FP}}}}
\end{eqnarray}
where the gauge-fixing action (coming from the Gaussian weight factor and the functional determinant) and the ghost action (coming from the inverse of the functional determinant) are (see appendix \ref{ghost})
\begin{eqnarray}
S_{\mbox{{\tiny GF}}} &=& \int d^8z\;\left(-\frac{1}{64g^2\xi C_2(A)}\mbox{Tr}V\{D^2,\bar{D}^2\}V-\overline{q}_iV\Phi_i-\overline{\Phi}_iVq_i-2g^2\xi(\overline{q}_iT^a\Phi_i)\frac{1}{\partial^2}(\overline{\Phi}_iT^aq_i)\right)\\
S_{\mbox{{\tiny FP}}} &=& \int d^8z\;\left(\frac{1}{C_2(A)}\mbox{Tr}\left[(c'+\overline{c}')(\mathcal{L}_{V/2}[(c+\overline{c})+\coth(\mathcal{L}_{V/2})(c-\overline{c})])\right]\right.\\
 && \left.-2g^2\xi\left[(\overline{q}_i+\overline{\Phi}_i)\overline{c}\left(\frac{1}{\partial^2}c'\right)q_i+\overline{q}_i\left(\frac{1}{\partial^2}\overline{c}'\right)c(q_i+\Phi_i)\right]\right).\nonumber
\end{eqnarray}
The gauge-fixing action generates the terms needed to cancel the quark superfield/vector superfield cross-terms.  However, in both the gauge-fixing and ghost actions, the gauge-fixing term also leads to non-local terms as stated above.  Most of the non-local terms consist of only two fields (two quark or two ghost superfields) and will therefore modify the propagators, in this case by generating mass terms.  The non-local terms consisting of more than two fields are at first sight problematic.  Only two vertices are of this kind, corresponding to interactions between one quark superfield and two ghost superfields.  Their effects will be investigated in the next section, after the propagators and vertices are obtained.

The free and interacting parts of the actions are easily found by expansion.  For the vector superfield, the free action can be simplified using projection operators (see appendix \ref{notation})
\begin{eqnarray}
S_V^0 &=& \int d^8z\;\left(V^a\left[\frac{1}{64g^2}\left(D^\alpha\bar{D}^2D_\alpha+\bar{D}_{\dot{\alpha}}D^2\bar{D}^{\dot{\alpha}}-\frac{1}{\xi}\{D^2,\bar{D}^2\}\right)\delta^{ab}+\frac{\mathcal{M}^{2ab}}{4g^2}\right]V^b+\overline{q}_iVq_i\right)\nonumber\\
 &=& \frac{1}{2}\int d^8z\;\left(V^a\left[-\frac{1}{2g^2}\left(P_T+\frac{1}{\xi}P_0-\frac{\mathcal{M}^2}{\partial^2}\right)^{ab}\partial^2\right]V^b\right)+\int d^8z\;\overline{q}_iVq_i
\end{eqnarray}
which gives the propagator
\begin{equation}
\bra{0}T\{V^a(z_1)V^b(z_2)\}\ket{0}=-2ig^2\left[\left(\frac{1}{\partial_1^2-\mathcal{M}^2}\right)^{ab}P_T+\xi\left(\frac{1}{\partial_1^2-\xi\mathcal{M}^2}\right)^{ab}P_0\right]\delta_{12}.
\end{equation}
The quark superfield free action is simply
\begin{eqnarray}
S_\Phi^0 &=& \int d^8z\;\left(\overline{\Phi}_i\Phi_i-2g^2\xi(\overline{q}_iT^a\Phi_i)\frac{1}{\partial^2}(\overline{\Phi}_iT^aq_i)\right)\nonumber\\
 &=& \frac{1}{2}\int d^8z\;\left(\overline{\Phi}_{in}\left[\delta_{ij}\delta_{nm}-\xi \frac{M_{in,jm}^2}{\partial^2}\right]\Phi_{jm}+\Phi_{in}\left[\delta_{ij}\delta_{nm}-\xi \frac{M_{jm,in}^2}{\partial^2}\right]\overline{\Phi}_{jm}\right)
\end{eqnarray}
and the free propagator becomes (notice that since the superpotential is zero no free propogators between $\Phi$ and $\Phi$ or between $\overline{\Phi}$ and $\overline{\Phi}$ appear)
\begin{equation}
\bra{0}T\{\Phi_{in}(z_1)\overline{\Phi}_{jm}(z_2)\}\ket{0}=i\left(\frac{\partial_1^2}{\partial_1^2-\xi M^2}\right)_{in,jm}P_2\delta_{12}=i\left[\delta_{ij}\delta_{nm}+\xi M_{in,jm}^{2ab}\left(\frac{1}{\partial_1^2-\xi\mathcal{M}^2}\right)^{ab}\right]P_2\delta_{12}.
\end{equation}
Finally the ghost superfield free action is
\begin{eqnarray}
S_g^0 &=& \int d^8z\;\left(\frac{1}{k}\mbox{Tr}\left[\overline{c}'c-c'\overline{c}\right]-2g^2\xi\left[\overline{q}_i\left(\frac{1}{\partial^2}\overline{c}\right)c'q_i+\overline{q}_i\overline{c}'\left(\frac{1}{\partial^2}c\right)q_i\right]\right)\nonumber\\
 &=& \int d^8z\;\left(\overline{c}'^a\left[\delta^{ab}-\xi\frac{\mathcal{M}^{2ab}}{\partial^2}\right]c^b-c'^a\left[\delta^{ab}-\xi\frac{\mathcal{M}^{2ba}}{\partial^2}\right]\overline{c}^b\right)
\end{eqnarray}
leading to the propagators
\begin{eqnarray}
\bra{0}T\{c^a(z_1)\overline{c}'^b(z_2)\}\ket{0} &=& i\left(\frac{\partial_1^2}{\partial_1^2-\xi\mathcal{M}^2}\right)^{ab}P_2\delta_{12}\\
\bra{0}T\{\overline{c}^a(z_1)c'^b(z_2)\}\ket{0} &=& -i\left(\frac{\partial_1^2}{\partial_1^2-\xi\mathcal{M}^2}\right)^{ba}P_1\delta_{12}.
\end{eqnarray}
Here $M_{in,jm}^{2ab}=2g^2(\overline{q}_jT^b)_m(T^aq_i)_n$ and the vector and ghost superfield mass matrix is $\mathcal{M}^{2ab}=\sum M_{in,in}^{2ab}$ while the quark superfield mass matrix is $M_{in,jm}^2=\sum M_{in,jm}^{2aa}$.  As pointed out before all non-local terms involving exactly two superfields modify the free propagators by generating mass terms.  This occurs since the projection operators $\{P_T,P_1,P_2\}$ of the free propagators absorb the extra $\frac{1}{\partial^2}$ factor of these non-local terms to produce the corresponding mass terms.  Therefore the only non-local terms left are in the interacting actions and involve one quark and two ghost superfields
\begin{eqnarray}
S_V^{\mbox{{\tiny int}}} &=& \frac{1}{64g^2C_2(A)}\mbox{Tr}\left[\int d^8z\;\left(\bar{D}^2D^\alpha V[V,D_\alpha V]-\frac{1}{4}[V,D^\alpha V]\bar{D}^2[V,D_\alpha V]\right.\right.\\
 && \left.\left.-\frac{1}{3}\bar{D}^2D^\alpha V[V,[V,D_\alpha V]]+\cdots\right)+h.c.\right]+\int d^8z\;\overline{q}_i\left[\frac{V^3}{3!}+\frac{V^4}{4!}+\cdots\right]q_i\nonumber\\
S_\Phi^{\mbox{{\tiny int}}} &=& \int d^8z\;\left(\overline{q}_i\left[\frac{V^2}{2}+\cdots\right]\Phi_i+\overline{\Phi}_i\left[\frac{V^2}{2}+\cdots\right]q_i+\overline{\Phi}_i\left[V+\frac{V^2}{2}+\cdots\right]\Phi_i\right)\\
S_g^{\mbox{{\tiny int}}} &=& \frac{1}{C_2(A)}\mbox{Tr}\int d^8z\;\left(\frac{1}{2}(c'+\overline{c}')[V,c-\overline{c}]+\frac{1}{12}(c'+\overline{c}')[V,[V,c-\overline{c}]]\cdots\right)\label{ghostint}\\
 && -2g^2\xi\int d^8z\;\left[\overline{\Phi}_i\overline{c}\left(\frac{1}{\partial^2}c'\right)q_i+\overline{q}_i\left(\frac{1}{\partial^2}\overline{c}'\right)c\Phi_i\right].\nonumber
\end{eqnarray}
Notice that, apart from non-locality issues, the Higgs mechanism in SUSY theories is similar to the Higgs mechanism in non-SUSY theories.  It leads to gauge-dependent mass terms for quark and ghost superfields and to quark superfield/ghost superfield/ghost superfield interactions as in non-SUSY theories.  In addition notice that all non-local terms disappear in superLorentz gauge ($\xi=0$).  Consequently one can undertake all computations in this specific gauge without worrying about non-locality.  The next section is devoted to show that the non-local vertices are well-behaved in the effective action at one-loop in any gauge.  The propagators and vertices in the GRS formalism \cite{Grisaru:1979wc} are given in appendix \ref{propagatorvertex}.

\section{Non-local terms in the effective action at one-loop}\label{oneloop}

The goal here is to compute one-loop contributions to the effective action coming from non-local terms in general gauge.  The interest lies in terms that could spoil the locality of the theory at one-loop.  By inspection the only possible divergent diagrams involving non-local vertices can be grouped according to their external superfields (here the zero superpotential decreases greatly the number of diagrams).

\begin{figure}[ht]\begin{center}\includegraphics[scale=1]{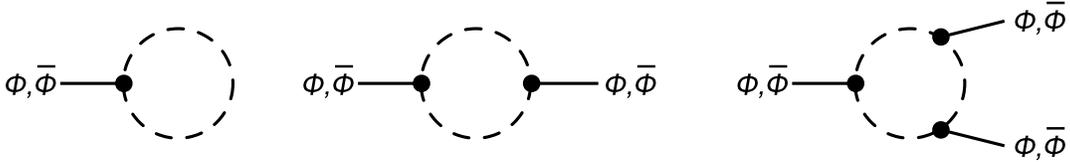}
\caption{Diagrams renormalizing the superpotential.
\label{Fig:Superpotential}}
\end{center}\end{figure}

The first group shown in figure \ref{Fig:Superpotential} corresponds to diagrams renormalizing the superpotential.  They are all exactly zero by the chirality properties of the external superfields, as anticipated from the non-renormalization theorem of SUSY theories.  For example, in the case of external chiral quark superfields $\Phi(z)$, after integrating by parts all covariant derivatives on one $\delta$-function and on quark chiral superfields, one ends up with integrals of chiral superfields with projection operators over full superspace.  These simplify ($P_T\Phi(z)=0$, $P_1\Phi(z)=0$ and $P_2\Phi(z)=\Phi(z)$) and give integrals of naked chiral superfields over full superspace, which are identically zero.  The same is true of anti-chiral quark superfields $\overline{\Phi}(z)$ with $P_2$ replaced by $P_1$.  Consequently no superpotential is generated, as expected in perturbation theory of SUSY theories and the non-local vertices do not affect the theory at this level.

\begin{figure}[ht]\begin{center}\includegraphics[scale=1]{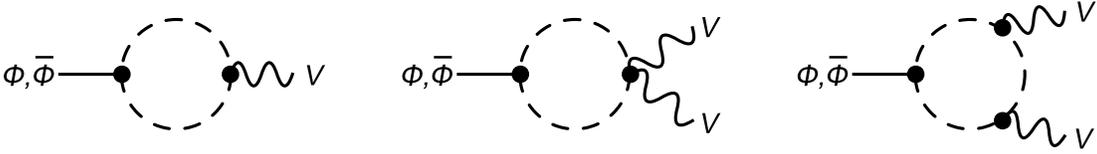}
\caption{Diagrams renormalizing the interactions between one quark superfield and any number of vector superfields.
\label{Fig:Phi1-V}}
\end{center}\end{figure}

\begin{figure}[ht]\begin{center}\includegraphics[scale=1]{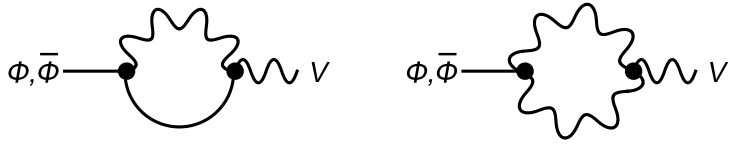}
\caption{Other relevant diagrams which renormalize the interactions between one quark superfield and one vector superfield.
\label{Fig:Corresponding-diagrams}}
\end{center}\end{figure}

The second group of diagrams of figure \ref{Fig:Phi1-V} renormalizes the interactions between one quark superfield and any number of vector superfields.  The number of external vector superfields is arbitrary since vector superfields have mass dimension zero in SUSY.  By gauge invariance all the diagrams in this group lead to the same infinite contributions.  For example, the first diagram of figure \ref{Fig:Phi1-V} with external chiral quark superfield $\Phi$ gives
\begin{multline}
2\times\frac{i^2}{2}\int d^8z_1d^8z_2\;\left\langle\frac{1}{2C_2(A)}\mbox{Tr}\overline{c}'(z_1)[V(z_1),c(z_1)](-2g^2\xi)\overline{q}_i\left(\frac{1}{\partial_2^2}\overline{c}'(z_2)\right)c(z_2)\Phi_i(z_2)\right\rangle\\
=-ig^2\xi f^{abc}(\overline{q}_iT^dT^e)_n\int\frac{d^4p}{(2\pi)^4}\frac{d^4k}{(2\pi)^4}d^4\theta\;V^a(-p,\theta)\left(\frac{1}{(p+k)^2+\xi\mathcal{M}^2}\right)^{cd}\left(\frac{1}{k^2+\xi\mathcal{M}^2}\right)^{eb}\Phi_{in}(p,\theta)\\
=-ig^2\xi f^{abc}(\overline{q}_iT^cT^b)_n\int\frac{d^4p}{(2\pi)^4}\frac{d^4k}{(2\pi)^4}d^4\theta\;V^a(-p,\theta)\frac{1}{k^2(p+k)^2}\Phi_{in}(p,\theta)+\mbox{finite}.
\end{multline}
With respect to non-SUSY theories, this diagram is equivalent to its non-SUSY analog since
\begin{equation}
\int d^4\theta\;V(\theta)\Phi(\theta)\supset\int d^4\theta\;(-\theta\sigma^\mu\bar{\theta}A_\mu)(i\theta\sigma^\nu\bar{\theta}\partial_\nu\phi).
\end{equation}
It also fulfills the same goal, i.e. it cancels gauge-dependent terms in figure \ref{Fig:Corresponding-diagrams}.  Moreover it is only logarithmically divergent as expected in SUSY theories.  This divergence has the same form as the field strength 
renormalization divergence thus it should be taken care off by the same counterterm.  Therefore the theory seems unaffected by non-locality issues for this group of diagrams.

\begin{figure}[ht]\begin{center}\includegraphics[scale=1]{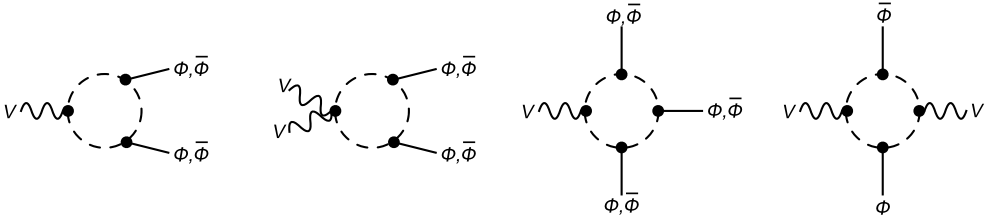}
\caption{Diagrams renormalizing the interactions between two or more quark superfields and any number of vector superfields.
\label{Fig:Phi2or3-V}}
\end{center}\end{figure}

The third group (see figure \ref{Fig:Phi2or3-V}) consists of diagrams with two or three external quark superfields (chiral or anti-chiral) and any number of vector superfields.  A simple computation shows that these diagrams are all finite and thus do not spoil the theory.  Indeed, since quark superfields and quark vacuum expectation values have mass dimension one, these diagrams have to be finite by dimensional analysis.

\begin{figure}[ht]\begin{center}\includegraphics[scale=1.3]{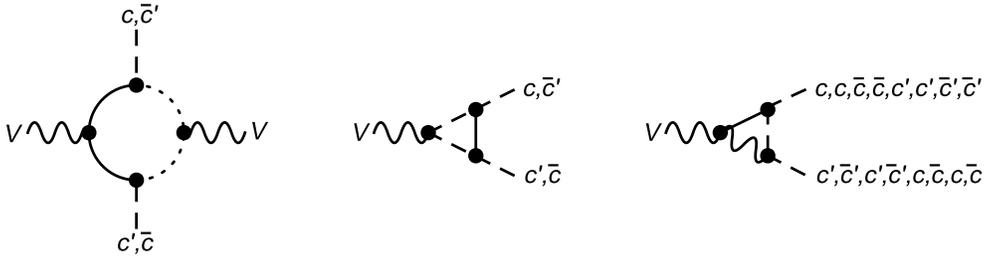}
\caption{Diagrams involving external ghost superfields.
\label{Fig:Unphysical-processes}}
\end{center}\end{figure}

The fourth and last group of diagrams of figure \ref{Fig:Unphysical-processes} is defined by unphysical processes where ghost superfields appear on external legs.  These diagrams are the most dangerous since the non-locality may lie on the external legs.  However, diagrams with non-local factors on external legs become local since the non-local factors disappear in the integration process.  Indeed, integration by parts pushes the appropriate covariant derivatives on the external ghost superfields with a $\frac{1}{\partial^2}$ factor which gives rise to the appropriate projection operators.  By the chirality properties of the ghost superfields, the non-local factor then disappears.  For example, the second diagram of \ref{Fig:Unphysical-processes} with external chiral ghost superfields $c$ and $c'$ contains
\begin{multline}
\int d^8z_1d^8z_2d^8z_3\;\left(\frac{\partial_2^2}{\partial_2^2-\xi M^2}\right)_{in,jm}P_2\delta_{12}^8\left(\frac{\partial_1^2}{\partial_1^2-\xi\mathcal{M}^2}\right)^{ab}P_1\delta_{13}^8\\
\;\;\;\;\;\;\;\;\;\;\;\;\;\;\;\;\;\;\;\;\;\;\;\;\;\;\;\;\;\;\;\;\;\;\;\;\;\;\;\;\;\;\times\left(\frac{\partial_2^2}{\partial_2^2-\xi\mathcal{M}^2}\right)^{cd}P_1\delta_{23}^8\frac{1}{\partial_1^2}c'^e(z_1)c^f(z_2)V^g(z_3)\\
=\int d^8z_1d^8z_2d^8z_3\;\left(\frac{\partial_2^2}{\partial_2^2-\xi M^2}\right)_{in,jm}P_2\delta_{12}^8\left(\frac{1}{\partial_1^2-\xi\mathcal{M}^2}\right)^{ab}\delta_{13}^8\\
\;\;\;\;\;\;\;\;\;\;\;\;\;\;\;\;\;\;\;\;\;\;\;\;\;\;\;\;\;\;\;\;\;\;\;\;\;\;\;\;\;\;\times\left(\frac{\partial_2^2}{\partial_2^2-\xi\mathcal{M}^2}\right)^{cd}P_1\delta_{23}^8\frac{\bar{D}_1^2D_1^2}{16\partial_1^2}c'^e(z_1)c^f(z_2)V^g(z_3)\\
\end{multline}
where $P_1$ is naturally integrated by parts on the ghost superfield $c'^e(z_1)$ leading to $P_2c'^e(z_1)=c'^e(z_1)$.  Moreover, by dimensional analysis these diagrams are all finite.

From this analysis the theory thus seems well-defined at one-loop in any gauge.  Moreover, by similar considerations one expects the theory to be well-defined at any order in perturbation theory.  In fact, in physical processes ghosts never occur as external fields and thus have to be contracted.  This helps the analysis since ghost propagators carry an extra $\partial^2$ factor in their numerator which cancels the non-local $\frac{1}{\partial^2}$ contributions of the vertices.  These diagrams should then have a clear meaning.  For unphysical processes with external ghost superfields the $\frac{1}{\partial^2}$ factor of non-local vertices is taken care of by covariant derivatives and the chiral properties of the ghost superfields.  In the end non-local effects only seem to generate less divergent quantum corrections and as a result additional counterterms are not required.  These reasons suggest that spontaneously broken $SU(N_c)$ superQCD with matter is well-defined in general gauge in perturbation theory.

\section{Conclusion}\label{Conclusion}

Supersymmetric $R_\xi$-gauge for super Yang-Mills theory with spontaneously broken gauge group leads to subtleties which deserve investigation.  The chiral choice of the gauge-fixing term introduces non-local terms which could spoil the locality of the theory.  It is shown here that these terms do not threaten the consistency of the theory at one-loop.  In fact, it parallels quite closely the non-SUSY case.  Indeed, non-zero quark vacuum expectation values lead to vector mass terms by the Higgs mechanism and part of the newly introduced non-local terms give rise to gauge-dependent mass terms for quark and ghost superfields.  Moreover the remaining non-local vertices result in analogous non-SUSY quantum corrections.  The non-renormalization theorem forces some of the corrections related to non-locality to be exactly zero while the non-zero diagrams left over are at worst logarithmically divergent and cancel gauge-dependent terms in well-behaved diagrams.  No additional counterterms seems to be required.  Adding a non-zero superpotential brings more free propagators but the general idea stays the same.  Simplified computations can be performed in the superLorentz gauge where all problematic non-local terms disappear and the theory gives expected results for the $\beta$-function of the gauge coupling.  Unfortunately the computation is long and tedious and won't be reported here.  Other choices of gauge groups do not seem to complicate the problem further.

\section*{Acknowledgment}\label{Acknowledgment}

The author would like to thank T. Banks, E. Andriyash, M. Dine and S. Thomas for useful discussions.  The author would also like to thank W. Siegel and P. van Nieuwenhuizen for pointing out references.

\appendix

\section{Notation}\label{notation}

The notation conventions used throughout the paper (see \cite{Ovrut:1981wa}) are reported here.  The group generators $T^a$ in the representation $R$ are chosen hermitian and satisfy the following identities
\begin{eqnarray}
[T^a,T^b] &=& if^{abc}T^c\\
\mbox{Tr}(T^aT^b) &=& T(R)\delta^{ab}\\
(T^aT^a)_{nm} &=& C_2(R)\delta_{nm}\\
f^{acd}f^{bcd} &=& C_2(A)\delta^{ab}\\
T(A) &=& C_2(A).\\
\end{eqnarray}
$f^{abc}$ are the structure constants and $T(R),C_2(R)$ are the Casimir coefficients of the representation $R$.  In superspace the compact notation is $\delta_{12}=\delta^8(z_1-z_2)=\delta^4(x_1-x_2)\delta^4(\theta_1-\theta_2)=\delta_{12}^x\delta_{12}^\theta$.  Several useful identities for integrals over superspace are (some make sense only in integrals)
\begin{eqnarray}
D_1^\alpha\delta_{12} &=& -D_2^\alpha\delta_{12}\\
D_1^2\delta_{12} &=& D_2^2\delta_{12}\\
\delta_{12}^\theta \bar{D}_1^2D_1^2\delta_{12} &=& \delta_{12}^\theta \bar{D}_2^2D_2^2\delta_{12}=16\delta_{12}\\
\delta_{12}^\theta D_1^2\bar{D}_1^2\delta_{12} &=& \delta_{12}^\theta D_2^2\bar{D}_2^2\delta_{12}=16\delta_{12}\\
\delta_{12}^\theta D_1^\alpha\bar{D}_1^2D_{1\alpha}\delta_{12} &=& \delta_{12}^\theta D_2^\alpha\bar{D}_2^2D_{2\alpha}\delta_{12}=16\delta_{12}\\
D^2\bar{D}^2D^2 &=& 16\partial^2D^2\\
\bar{D}^2D^2\bar{D}^2 &=& 16\partial^2\bar{D}^2\\
\{D_\alpha,\bar{D}_{\dot{\alpha}}\} &=& -2i\sigma_{\alpha\dot{\alpha}}^\mu\partial_\mu\\
\sigma_{\alpha\dot{\alpha}}^\mu\bar{\sigma}_\nu^{\dot{\alpha}\alpha} &=& -2g^{\mu\nu}\\
\left.\right.[D_\alpha,\bar{D}^2] &=& -4i\sigma_{\alpha\dot{\alpha}}^\mu\partial_\mu\bar{D}^{\dot{\alpha}}\\
\left.\right.[\bar{D}_{\dot{\alpha}},D^2] &=& 4i\sigma_{\alpha\dot{\alpha}}^\mu\partial_\mu D^\alpha\\
D_\alpha D_\beta &=& \frac{1}{2}\epsilon_{\alpha\beta}D^2\\
\bar{D}_{\dot{\alpha}}\bar{D}_{\dot{\beta}} &=& -\frac{1}{2}\epsilon_{\dot{\alpha}\dot{\beta}}\bar{D}^2.
\end{eqnarray}
From the $\delta_{12}^\theta$-function reduction formulae (A.5-7), one can focus only on integrals with naked $\delta$-functions and one $\delta$-function with four covariant derivatives.
\noindent
One also introduces the projection operators $P_i=\{P_1,P_2,P_T\}$,
\begin{equation}
P_1=\frac{D^2\bar{D}^2}{16\partial^2}\;\;\;\;\;\;P_2=\frac{\bar{D}^2D^2}{16\partial^2}\;\;\;\;\;\;P_0=P_1+P_2
\end{equation}
and
\begin{equation}
P_T=-\frac{D^\alpha\bar{D}^2D_\alpha}{8\partial^2}=-\frac{\bar{D}_{\dot{\alpha}}D^2\bar{D}^{\dot{\alpha}}}{8\partial^2}.
\end{equation}
As their name implies they obey the following relations
\begin{equation}
\sum_{i=\{1,2,T\}}P_i=1\;\;\;\;\;\;P_iP_j=\delta_{ij}P_j.
\end{equation}
Moreover chiral superfields $\Phi(z)$ obey
\begin{equation}
P_T\Phi(z)=0\;\;\;\;\;\;P_1\Phi(z)=0\;\;\;\;\;\;P_2\Phi(z)=\Phi(z).
\end{equation}
Two additional operators given by
\begin{equation}
P_+=\frac{D^2}{4(\partial^2)^{\frac{1}{2}}}\;\;\;\;\;\;P_-=\frac{\bar{D}^2}{4(\partial^2)^{\frac{1}{2}}}
\end{equation}
are helpful in inverting matrix with covariant derivatives (see \cite{Ovrut:1981wa}).  Those are useful when one has a non-zero superpotential which mixes chiral quark superfields together.
\noindent
Finally Fourier transforms are defined as
\begin{equation}
A(x,\theta)=\int\frac{d^4k}{(2\pi)^4}\;A(k,\theta)e^{-ik\cdot x}
\end{equation}
and integrals over half superspace are converted into integrals over full superspace with the help of
\begin{equation}
\int d^4xd^2\theta\;\left(-\frac{1}{4}\bar{D}^2\right)F=\int d^4xd^2\theta d^2\bar{\theta}\;F.
\end{equation}
This is possible since derivatives in superspace are the same than integrals.

\section{Ghost action}\label{ghost}

The ghost action is found by usual techniques.  Using integral representations of $\delta$-functions the functional determinant can be written as
\begin{eqnarray}
\Delta(V) &=& \int\mathbbm{D}\Lambda\mathbbm{D}\overline{\Lambda}\;\delta[F(V^{\Lambda,\overline{\Lambda}})-f]\delta[\overline{F}(V^{\Lambda,\overline{\Lambda}})-\overline{f}]\nonumber\\
 &=& \int\mathbbm{D}\Lambda\mathbbm{D}\overline{\Lambda}\mathbbm{D}\Lambda'\mathbbm{D}\overline{\Lambda}'\;\exp\left(\int d^8z\left[\Lambda'^a\left(\frac{\delta F}{\delta\Lambda}\Lambda+\frac{\delta F}{\delta\overline{\Lambda}}\overline{\Lambda}\right)^a+\overline{\Lambda}'^a\left(\frac{\delta\overline{F}}{\delta\Lambda}\Lambda+\frac{\delta\overline{F}}{\delta\overline{\Lambda}}\overline{\Lambda}\right)^a\right]\right)
\end{eqnarray}
where $\Lambda'$ and $\overline{\Lambda}'$ are general superfields\symbolfootnote[1]{Unlike the gauge parameters $\Lambda$ and $\overline{\Lambda}$ which are chiral and anti-chiral superfields respectively.} and the derivatives are to first order in the gauge parameters.  From the field transformation properties
\begin{eqnarray}
Q_i &\rightarrow& Q_i'=e^{-i\Lambda}Q_i\nonumber\\
e^V &\rightarrow& e^{V'}=e^{-i\overline{\Lambda}}e^{V}e^{i\Lambda}
\end{eqnarray}
the variations and appropriate derivatives are easily found (here $\mathcal{L}_XY=[X,Y]$)
\begin{eqnarray}
\Delta(V) &=& \int\mathbbm{D}\Lambda\mathbbm{D}\overline{\Lambda}\mathbbm{D}\Lambda'\mathbbm{D}\overline{\Lambda}'\;\exp\left(i\int d^8z\;\left[\Lambda'^a\bar{D}^2(\mathcal{L}_{V/2}[(\Lambda+\overline{\Lambda})+\coth(\mathcal{L}_{V/2})(\Lambda-\overline{\Lambda})])^a\right.\right.\nonumber\\
 && \left.\left.+\overline{\Lambda}'^aD^2(\mathcal{L}_{V/2}[(\Lambda+\overline{\Lambda})+\coth(\mathcal{L}_{V/2})(\Lambda-\overline{\Lambda})])^a\right.\right.\\
 && \left.\left.+\Lambda'^a\bar{D}^2\frac{2g^2\xi}{\partial^2}(\overline{q}_i+\overline{\Phi}_i)\overline{\Lambda}T^aq_i-\overline{\Lambda}'^aD^2\frac{2g^2\xi}{\partial^2}\overline{q}_iT^a\Lambda (q_i+\Phi_i)\right]\right).\nonumber
\end{eqnarray}
To invert it, one uses anti-commuting superfields $b^a$, $\overline{b}^a$, $c^a$ and $\overline{c}^a$ instead of commuting superfields $\Lambda'^a$, $\overline{\Lambda}'^a$, $\Lambda^a$ and $\overline{\Lambda}^a$ respectively, which gives
\begin{eqnarray}
\Delta^{-1}(V) &=& \int\mathbbm{D}c\mathbbm{D}\overline{c}\mathbbm{D}b\mathbbm{D}\overline{b}\;\exp\left(i\int d^8z\;\left[b^a\bar{D}^2(\mathcal{L}_{V/2}[(c+\overline{c})+\coth(\mathcal{L}_{V/2})(c-\overline{c})])^a\right.\right.\nonumber\\
 && \left.\left.+\overline{b}^aD^2(\mathcal{L}_{V/2}[(c+\overline{c})+\coth(\mathcal{L}_{V/2})(c-\overline{c})])^a\right.\right.\nonumber\\
 && \left.\left.+b^a\bar{D}^2\frac{2g^2\xi}{\partial^2}(\overline{q}_i+\overline{\Phi}_i)\overline{c}T^aq_i-\overline{b}^aD^2\frac{2g^2\xi}{\partial^2}\overline{q}_iT^ac(q_i+\Phi_i)\right]\right)\\
 &=& \int\mathbbm{D}c\mathbbm{D}\overline{c}\mathbbm{D}c'\mathbbm{D}\overline{c}'\;e^{iS_{\mbox{{\tiny FP}}}}.\nonumber
\end{eqnarray}
Here integration by parts was used to write the general anti-commuting superfields $b$ and $\overline{b}$ as chiral and anti-chiral anti-commuting superfields $c'=\bar{D}^2b$ and $\overline{c}'=D^2\overline{b}$.  The ability to write the ghost action only in terms of chiral and anti-chiral ghost superfields was expected since the gauge-fixing term is chiral.  The ghost action is
\begin{eqnarray}
S_{\mbox{{\tiny FP}}} &=& \int d^8z\;\left(\frac{1}{C_2(A)}\mbox{Tr}\left[(c'+\overline{c}')(\mathcal{L}_{V/2}[(c+\overline{c})+\coth(\mathcal{L}_{V/2})(c-\overline{c})])\right]\right.\nonumber\\
 && \left.-2g^2\xi\left[\left(\frac{1}{\partial^2}\left[(\overline{q}_i+\overline{\Phi}_i)\overline{c}\right]\right)c'q_i+\overline{q}_i\overline{c}'\left(\frac{1}{\partial^2}\left[c(q_i+\Phi_i)\right]\right)\right]\right)
\end{eqnarray}
where the generators in the first term are chosen to be in the adjoint representation $A$ of the gauge group.  Notice again the presence of non-local terms $\frac{1}{\partial^2}$ in $S_{\mbox{{\tiny FP}}}$, as for $S_{\mbox{{\tiny GF}}}$.

\section{Propagators and vertices in GRS formalism}\label{propagatorvertex}

This appendix lists the free propagators and vertices in the GRS formalism \cite{Grisaru:1979wc}.  The propagators are
\begin{equation}
\bra{0}T\{V^a(z_1)V^b(z_2)\}\ket{0}_{\mbox{{\tiny GRS}}}=-2ig^2\left[\left(\frac{1}{\partial_1^2-\mathcal{M}^2}\right)^{ab}P_T+\xi\left(\frac{1}{\partial_1^2-\xi\mathcal{M}^2}\right)^{ab}P_0\right]\delta_{12}
\end{equation}
\begin{equation}
\bra{0}T\{\Phi_{in}(z_1)\overline{\Phi}_{jm}(z_2)\}\ket{0}_{\mbox{{\tiny GRS}}}=i\left[\delta_{ij}\delta_{nm}+\xi M_{in,jm}^{2ab}\left(\frac{1}{\partial_1^2-\xi\mathcal{M}^2}\right)^{ab}\right]\delta_{12}
\end{equation}
\begin{equation}
\bra{0}T\{c^a(z_1)\overline{c}'^b(z_2)\}\ket{0}_{\mbox{{\tiny GRS}}}=i\left(\frac{\partial_1^2}{\partial_1^2-\xi\mathcal{M}^2}\right)^{ab}\delta_{12}
\end{equation}
\begin{equation}
\bra{0}T\{\overline{c}^a(z_1)c'^b(z_2)\}\ket{0}_{\mbox{{\tiny GRS}}}=-i\left(\frac{\partial_1^2}{\partial_1^2-\xi\mathcal{M}^{2*}}\right)^{ab}\delta_{12}
\end{equation}
with $\mathcal{M}^{2*ab}=\mathcal{M}^{2ba}$.  The GRS propagators are easily obtained by removing the $P_{1,2}$ projectors of chiral and anti-chiral field propagators, i.e. quark and ghost propagators (the vector superfield free propagator stays the same).  The vertices are obtained in the same way from the interaction action.  Vertices involving vector superfields only are
\begin{fmffile}{VVV-Vertex}
\begin{eqnarray}
\begin{fmfgraph*}(20,10)
\fmfleftn{i}{2}
\fmfrightn{o}{1}
\fmflabel{$V^a(z_1)$}{i1}
\fmflabel{$V^b(z_2)$}{i2}
\fmflabel{$V^c(z_3)$}{o1}
\fmf{photon}{i1,v1}
\fmf{photon}{i2,v1}
\fmf{photon}{v1,o1}
\fmfdot{v1}
\end{fmfgraph*} &=& \left.\frac{i\delta^3S}{\delta V^{a}(z_1)\delta V^{b}(z_2)\delta V^{c}(z_3)}\right|_0\nonumber\\
 &=& \int d^8z\;\left(\frac{-f^{a'b'c'}}{64g^2}\left[\bar{D}^2D^\alpha\delta^{ca'}\delta^8_{3z}D_\alpha\delta^{bb'}\delta^8_{2z}\delta^{ac'}\delta^8_{1z}+\bar{D}^2D^\alpha\delta^{ca'}\delta^8_{3z}D_\alpha\delta^{ab'}\delta^8_{1z}\delta^{bc'}\delta^8_{2z}\right.\right.\nonumber\\
 && \left.\left.+\bar{D}^2D^\alpha\delta^{ba'}\delta^8_{2z}D_\alpha\delta^{cb'}\delta^8_{3z}\delta^{ac'}\delta^8_{1z}+\bar{D}^2D^\alpha\delta^{aa'}\delta^8_{1z}D_\alpha\delta^{cb'}\delta^8_{3z}\delta^{bc'}\delta^8_{2z}\right.\right.\nonumber\\
 && \left.\left.+\bar{D}^2D^\alpha\delta^{ba'}\delta^8_{2z}D_\alpha\delta^{ab'}\delta^8_{1z}\delta^{cc'}\delta^8_{3z}+\bar{D}^2D^\alpha\delta^{aa'}\delta^8_{1z}D_\alpha\delta^{bb'}\delta^8_{2z}\delta^{cc'}\delta^8_{3z}+h.c.\right]\right.\nonumber\\
 && \left.+\frac{i(\overline{q}_iT^{a'}T^{b'}T^{c'}q_i)}{6}\left[\delta^{ca'}\delta^8_{3z}\delta^{bb'}\delta^8_{2z}\delta^{ac'}\delta^8_{1z}+\delta^{ca'}\delta^8_{3z}\delta^{ab'}\delta^8_{1z}\delta^{bc'}\delta^8_{2z}\right.\right.\nonumber\\
 && \left.\left.+\delta^{ba'}\delta^8_{2z}\delta^{cb'}\delta^8_{3z}\delta^{ac'}\delta^8_{1z}+\delta^{aa'}\delta^8_{1z}\delta^{cb'}\delta^8_{3z}\delta^{bc'}\delta^8_{2z}\right.\right.\nonumber\\
 && \left.\left.+\delta^{ba'}\delta^8_{2z}\delta^{ab'}\delta^8_{1z}\delta^{cc'}\delta^8_{3z}+\delta^{aa'}\delta^8_{1z}\delta^{bb'}\delta^8_{2z}\delta^{cc'}\delta^8_{3z}\right]\right)\nonumber\\
 &=& \int d^8z\;\left(\frac{-f^{a'b'c'}}{64g^2}\left[\bar{D}^2D^\alpha\delta^{aa'}\delta^8_{1z}D_\alpha\delta^{bb'}\delta^8_{2z}\delta^{cc'}\delta^8_{3z}+\mbox{permutations}+h.c.\right]\right.\nonumber\\
 && \left.+\frac{i(\overline{q}_iT^{a'}T^{b'}T^{c'}q_i)}{6}\left[\delta^{aa'}\delta^8_{1z}\delta^{bb'}\delta^8_{2z}\delta^{cc'}\delta^8_{3z}+\mbox{permutations}\right]\right)
\end{eqnarray}
\end{fmffile}
\begin{fmffile}{VVVV-Vertex}
\begin{eqnarray}
\begin{fmfgraph*}(20,10)
\fmfleftn{i}{2}
\fmfrightn{o}{2}
\fmflabel{$V^a(z_1)$}{i1}
\fmflabel{$V^b(z_2)$}{i2}
\fmflabel{$V^c(z_3)$}{o1}
\fmflabel{$V^d(z_4)$}{o2}
\fmf{photon}{i1,v1}
\fmf{photon}{i2,v1}
\fmf{photon}{v1,o1}
\fmf{photon}{v1,o2}
\fmfdot{v1}
\end{fmfgraph*} &=& \int d^8z\;\left(\frac{if^{a'b'e'}f^{c'd'e'}}{64g^2}\left[\left(\frac{1}{4}\delta^{aa'}\delta^8_{1z}D^\alpha\delta^{bb'}\delta^8_{2z}\bar{D}^2[\delta^{cc'}\delta^8_{3z}D_\alpha\delta^{dd'}\delta^8_{4z}]\right.\right.\right.\nonumber\\
 && \left.\left.\left.+\frac{1}{3}\bar{D}^2D^\alpha\delta^{aa'}\delta^8_{1z}\delta^{bb'}\delta^8_{2z}\delta^{cc'}\delta^8_{3z}D_\alpha\delta^{dd'}\delta^8_{4z}\right)+\mbox{permutations}+h.c.\right]\right.\\
 && \left.+\frac{i(\overline{q}_iT^{a'}T^{b'}T^{c'}T^{d'}q_i)}{24}\left[\delta^{aa'}\delta^8_{1z}\delta^{bb'}\delta^8_{2z}\delta^{cc'}\delta^8_{3z}\delta^{dd'}\delta^8_{4z}+\mbox{permutations}\right]\right).\nonumber
\end{eqnarray}
\end{fmffile}
Vertices involving vector superfields and quark superfields are
\begin{fmffile}{QVV-Vertex}
\begin{eqnarray}
\begin{fmfgraph*}(20,10)
\fmfleftn{i}{2}
\fmfrightn{o}{1}
\fmflabel{$V^a(z_1)$}{i1}
\fmflabel{$V^b(z_2)$}{i2}
\fmflabel{$\Phi_{in}(z_3)$}{o1}
\fmf{photon}{i1,v1}
\fmf{photon}{i2,v1}
\fmf{plain}{v1,o1}
\fmfdot{v1}
\end{fmfgraph*} &=& \int d^8z\;\frac{i(\overline{q}_i\{T^a,T^b\})_n}{2}\delta^8_{1z}\delta^8_{2z}\delta^8_{3z}\\
\begin{fmfgraph*}(20,10)
\fmfleftn{i}{2}
\fmfrightn{o}{1}
\fmflabel{$V^a(z_2)$}{i1}
\fmflabel{$V^b(z_3)$}{i2}
\fmflabel{$\overline{\Phi}_{in}(z_1)$}{o1}
\fmf{photon}{i1,v1}
\fmf{photon}{i2,v1}
\fmf{plain}{v1,o1}
\fmfdot{v1}
\end{fmfgraph*} &=& \int d^8z\;\frac{i(\{T^a,T^b\}q_i)_n}{2}\delta^8_{1z}\delta^8_{2z}\delta^8_{3z}
\end{eqnarray}
\end{fmffile}
with
\begin{fmffile}{QQV-Vertex}
\begin{eqnarray}
\begin{fmfgraph*}(20,10)
\fmfleftn{i}{2}
\fmfrightn{o}{1}
\fmflabel{$\overline{\Phi}_{in}(z_1)$}{i1}
\fmflabel{$\Phi_{jm}(z_3)$}{i2}
\fmflabel{$V^a(z_2)$}{o1}
\fmf{plain}{i1,v1}
\fmf{plain}{i2,v1}
\fmf{photon}{v1,o1}
\fmfdot{v1}
\end{fmfgraph*} &=& \int d^8z\;i\delta_{ij}T_{nm}^a\delta^8_{1z}\delta^8_{2z}\delta^8_{3z}\\
\begin{fmfgraph*}(20,10)
\fmfleftn{i}{2}
\fmfrightn{o}{2}
\fmflabel{$\overline{\Phi}_{in}(z_1)$}{i1}
\fmflabel{$\Phi_{jm}(z_4)$}{i2}
\fmflabel{$V^a(z_2)$}{o1}
\fmflabel{$V^b(z_3)$}{o2}
\fmf{plain}{i1,v1}
\fmf{plain}{i2,v1}
\fmf{photon}{v1,o1}
\fmf{photon}{v1,o2}
\fmfdot{v1}
\end{fmfgraph*} &=& \int d^8z\;i\delta_{ij}\{T^a,T^b\}_{nm}\delta^8_{1z}\delta^8_{2z}\delta^8_{3z}\delta^8_{4z}.
\end{eqnarray}
\end{fmffile}
Vertices with vector superfields and ghost superfields consist of
\begin{fmffile}{ggV-Vertex}
\begin{eqnarray}
\begin{fmfgraph*}(20,10)
\fmfleftn{i}{2}
\fmfrightn{o}{1}
\fmflabel{$c'^a(z_1)$}{i1}
\fmflabel{$c^c(z_3)$}{i2}
\fmflabel{$V^b(z_2)$}{o1}
\fmf{dashes}{i1,v1}
\fmf{dashes}{i2,v1}
\fmf{photon}{v1,o1}
\fmfdot{v1}
\end{fmfgraph*} &=& \int d^8z\;\frac{-f^{abc}}{2}\delta^8_{1z}\delta^8_{2z}\delta^8_{3z}\label{ggVint}\\
\begin{fmfgraph*}(20,10)
\fmfleftn{i}{2}
\fmfrightn{o}{2}
\fmflabel{$\overline{c}'^a(z_1)$}{i1}
\fmflabel{$\overline{c}^d(z_4)$}{i2}
\fmflabel{$V^b(z_2)$}{o1}
\fmflabel{$V^c(z_3)$}{o2}
\fmf{dashes}{i1,v1}
\fmf{dashes}{i2,v1}
\fmf{photon}{v1,o1}
\fmf{photon}{v1,o2}
\fmfdot{v1}
\end{fmfgraph*} &=& \int d^8z\;\frac{\pm f^{abe}f^{cde}}{12}\delta^8_{1z}\delta^8_{2z}\delta^8_{3z}\delta^8_{4z}
\end{eqnarray}
\end{fmffile}
where the sign is related to equation (\ref{ghostint}).  Finally vertices involving chiral superfields and ghost superfields are
\begin{fmffile}{Qgg-Vertex}
\begin{eqnarray}
\begin{fmfgraph*}(20,10)
\fmfleftn{i}{2}
\fmfrightn{o}{1}
\fmflabel{$\overline{c}'^a(z_1)$}{i1}
\fmflabel{$c^b(z_2)$}{i2}
\fmflabel{$\Phi_{in}(z_3)$}{o1}
\fmf{dashes}{i1,v1}
\fmf{dashes}{i2,v1}
\fmf{plain}{v1,o1}
\fmfdot{v1}
\end{fmfgraph*} &=& \int d^8z\;(-2i)g^2\xi(\overline{q}_iT^aT^b)_n\left(\frac{1}{\partial^2}\delta^8_{1z}\right)\delta^8_{2z}\delta^8_{3z}\\
\begin{fmfgraph*}(20,10)
\fmfleftn{i}{2}
\fmfrightn{o}{1}
\fmflabel{$\overline{c}^a(z_2)$}{i1}
\fmflabel{$c'^b(z_3)$}{i2}
\fmflabel{$\overline{\Phi}_{in}(z_1)$}{o1}
\fmf{dashes}{i1,v1}
\fmf{dashes}{i2,v1}
\fmf{plain}{v1,o1}
\fmfdot{v1}
\end{fmfgraph*} &=& \int d^8z\;(-2i)g^2\xi(T^aT^bq_i)_n\delta^8_{1z}\delta^8_{2z}\left(\frac{1}{\partial^2}\delta^8_{3z}\right).
\end{eqnarray}
\end{fmffile}
These last two vertices are the only non-local vertices in general gauge.  All higher-order vertices (in $g$ after the rescaling $V\rightarrow2gV$) can easily be found from the interacting action.  In the GRS formalism, covariant derivatives are removed from the free propagators and therefore the vertices are easier to handle.  For example, vector superfield/ghost superfield/ghost superfield vertices in the GRS formalism are all the same \cite{Grisaru:1979wc} as shown in equation (\ref{ggVint}).  One must not forget to re-introduce appropriate covariant derivatives before computing diagrams.

\bibliographystyle{utcaps}
\bibliography{SUSYHiggs.bbl}

\end{document}